# Global Asteroid Risk Analysis

**Rumpf, Clemens**
University of Southampton, United Kingdom, C.Rumpf@soton.ac.uk

Potentially impacting asteroids were analysed for their impact risk on the Earth. To this end, the Asteroid Risk Mitigation Optimization and Research (ARMOR) tool is currently being developed. The tool's modules are described and their validation is documented. Based on the asteroid ephemeris, the tool calculates the impact location probability distribution on the surface of the Earth (in the literature, occasionally referred to as risk corridor). NASA's Near Earth Object (NEO) risk list served as the source for asteroid ephemerides. The Line of Variation (LOV) method was employed to find virtual impactors. While offering a simple and fast way of identifying virtual impactors, the method provides a low impactor identification rate. This is because the search space is tightly constricted to the LOV and thus excludes virtual impactors located elsewhere in the asteroid position uncertainty region. The method's performance was evaluated and suggestions for improvements are provided. Application of the tool showed that the asteroid threat is global in nature: impact locations were distributed widely over the Earth's surface. The global asteroid risk was estimated by combining the impact location probability distribution with Earth population data. The identification of high risk regions lead to a discussion about the dynamics of the risk associated with asteroids. The future response to an asteroid threat will depend on the risk that the asteroid poses to the population. The potential applicability of ARMOR as a decision-support tool for responding to the asteroid threat is outlined. The work is supported by the Marie Curie Initial Training Network Stardust, FP7-PEOPLE-2012-ITN, Grant Agreement 317185.

## I. INTRODUCTION

The asteroid threat concerns the entire world population. This is because any region on Earth is prone to experience an impact event[1,2]. Furthermore, asteroid impacts can release enough energy to affect extensive regions around the impact location, scaling up to global consequences and making every region vulnerable[3,4].

Albeit a low frequency event on human time scales[5,6], the potential destructive power of this natural disaster, as made tangible by the events over Tunguska[7] in 1908 and Chelyabinsk[8] in 2013, warrants development of response procedures and tools.

The asteroid threat distinguishes itself from other natural disasters in the sense that it is possible for humankind to turn an imminent impact event into a non-event by deflecting the asteroid. This capability only became feasible during the space age. The concept has already been demonstrated by the Deep Impact[9] mission that impacted a comet in 2005. With the realization that asteroid deflection is a technology very much in the realms of the possible, humankind has a moral obligation to actually develop technologies and implement procedures that will enable us to conduct a deflection mission. Space entities around the globe have realized this obligation and are cooperating to fulfil it. In 2013, the United Nations formally endorsed the recommended establishment of the Space Mission Planning Advisory Group (SMPAG) to lead the international effort to produce coherent response procedures and capabilities[10] in the future.

Currently, sky surveys such as Pan-STARRS and Catalina Sky Survey increase our knowledge about the asteroid environment in Earth's neighbourhood. More capable systems are planned to extend this knowledge base and also to focus on smaller asteroids (tens of metres diameter) of which less than 1% are discovered[11]. To be aware of an asteroid before it collides with the Earth is the first step in responding to the threat. As the observational capabilities improve, the discovery of an Earth-impacting asteroid becomes a matter of time. In the event of such a discovery, a response procedure needs to be in place that triggers action based on the asteroid characteristics (size, composition, trajectory), reaction time and the risk that the asteroid poses to the population and associated structures[12]. For small asteroids (maximum diameter of few tens of metres) and/or short reaction times (<3 years), evacuation of the threatened region might be the appropriate response[13]. For larger asteroids, and given enough reaction time, a deflection mission can be conceived[14]. In either case, the risk for people and assets on the ground will play a vital role in determining the appropriate response to the threat. If the risk is high and many people are at risk of losing their lives in an imminent impact, a proactive response is more likely to be initiated than in the case where the risk is low and no tangible loss is projected to occur during impact.

This work contributes to the risk assessment segment of the asteroid threat. NASA and ESA maintain freely accessible lists of asteroids that have a non-zero chance of impacting the Earth in the future[15,16]. The lists also provide the orbital elements of these asteroids. However, the potential impact locations of the asteroids are not available. Previous work by Bailey[17] provides the capability to quantify the damage and number of casualties caused by a specific asteroid if its impact location and properties are known. However, the tool lacks the capability to determine the impact location probability distribution associated with the orbital solution of a given asteroid. The research presented in this paper closes the gap between the information given in the risk lists and the tools that are able to determine impact effects. The impact location probability distribution of asteroids in the risk list is determined and a preliminary estimate of casualties is provided highlighting the global nature of the asteroid threat.

## II. METHOD

A software tool is under development that maps the asteroid ephemeris to an impact probability distribution on the surface of the Earth. The system is named Asteroid Risk Mitigation Optimization and Research (ARMOR) tool. Pending further development, three modules, written in Python, currently constitute the ARMOR tool and these are described in the following sections.

### II.I Solar System Propagator

The core of ARMOR is a solar system propagator. This module is a gravitational force model of the solar system in accordance with:

$$\ddot{r}_j = \sum_i \left[ -\frac{GM_i}{|r_{ij}|^3} r_{ij} \right] \qquad 1)$$

Where $i$ denotes all major bodies (Sun, planets and Pluto) and their moons in the solar system. The acceleration vector of the propagated body is $\ddot{r}_j$. The gravitational constant of each attracting body is $GM_i$ and $r_{ij}$ is the relative position vector from the attracting body to the propagated body.

The propagator computes the asteroid's trajectory through the solar system for a specified period of time, starting at a predefined initial ephemeris and epoch. The asteroid's initial ephemerides are obtained from the Jet Propulsion Laboratory's HORIZONS system[18]. The orbital data of the attracting bodies is retrieved from a lookup table. The trajectories of the asteroid and the other bodies are recorded for post processing such as close flyby detection.

### Propagator Validation

Since the solar system propagator's performance dictates critically the quality of all results obtained with ARMOR, the propagation accuracy was validated against ephemerides generated by HORIZONS. Fig. 1 shows the positional discrepancy between the two systems for asteroid 2006QV89. Positional discrepancy is the norm of the vector connecting the two asteroid positions at a given time as calculated by the propagator and HORIZONS. The asteroid 2006QV89 was chosen because it has eight close encounters with the Earth in the simulation period and therefore exhibits a highly perturbed orbit. A 15-year simulation period was chosen because this is the timeframe in which a deflection mission could be conducted. The root mean square position discrepancy over the simulation period is 0.0153 Earth radii (less than 100 km). The error tends to increase with longer simulation times. This behaviour can be expected for a propagator as the integration error accumulates over time. Considering the distances that describe the solar system (semi-major axis of Pluto is $5.87 * 10^9$ km), the result is highly accurate and gives confidence in future results that are based on the propagator's performance.

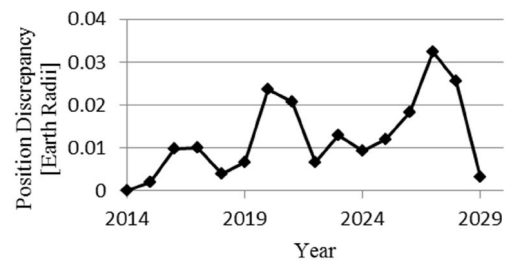

Fig. 1: Positional discrepancy for asteroid 2006QV89 between the propagator and HORIZONS for a 15 year simulation starting in 2014.

### II.II Virtual Impactor Finder

This module's purpose is to find the ephemeris variation of an asteroid that yields a collision with the Earth. Upon discovery of an asteroid, the most likely ephemeris for this asteroid is computed. This is called the nominal orbit solution and is the orbit solution that best fits the available measurements. However, the ephemeris has uncertainty associated with it. It is highly unlikely that an asteroid is discovered whose nominal orbit solution yields a future collision with the Earth. Instead, the locations within the orbit solution's uncertainty region that coincide with a future Earth collision are determined. Because of the way that asteroid position measurements are taken and processed, the uncertainty region stretches along the orbit of the asteroid and envelops a relatively thin region around the orbit. In other words, the geometry of the orbit is generally well known but the position of the asteroid on that orbit is less certain[19]. Fig. 2 is a depiction of this concept.

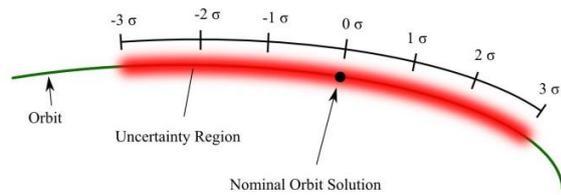

Fig. 2: Depiction of the uncertainty region of ephemeris in relation to the orbit.

The particular shape of the uncertainty region of the asteroid ephemeris gives way to the line of variation (LOV) method. The LOV is the direction in which the asteroid ephemeris shows the greatest uncertainty and it coincides with the asteroid's orbit. The method is an easy and fast way to sample the asteroid ephemeris to discover a variation that yields a future impact with the Earth. Effectively, the asteroid is positioned on different parts of the orbit in the vicinity of its nominal orbit solution. This means that sampling is done in a strictly one-dimensional way, constrained to the LOV. A repositioned asteroid is called a virtual asteroid and is propagated forward for impact detection. If no impact is detected, the epoch is varied again in the direction that minimizes the closest approach distance between the asteroid and the Earth. In other words, this simple search algorithm varies the nominal ephemeris towards the next local closest approach minimum. Eventually, the closest approach distance should become smaller than the radius of the Earth which corresponds to an impacting virtual asteroid called a virtual impactor. This method has been applied before[20] for asteroid 2011 AG5. Here, the method was applied to many more asteroids. The determined impact dates of the discovered virtual impactors (see Table I) show good agreement with the dates calculated by NASA and ESA. This finding serves as validation of the module.

II.III Risk Corridor Calculation

The Risk Corridor Calculation module determines the possible impact points of an asteroid on the surface of the Earth. The Virtual Impactor Finder module produces an ephemeris that corresponds to a first impact point. Starting at this ephemeris, the epoch is varied forwards and backwards slightly in time effectively creating more virtual asteroids to conduct impact analysis. A good step size for this variation is found to be 100 sec. If the new virtual asteroid is still on an impacting trajectory, the new impact location is recorded and the variation continues. Eventually, a new virtual asteroid will miss the Earth. In this case, the epoch variation step size is continuously decreased to 0.1 sec while the search algorithm finds the last virtual asteroid that just barely impacts Earth. This case is called a grazing case because the virtual asteroid is on a near tangential trajectory with respect to the surface of the Earth. Two grazing cases exist: one where the virtual asteroid just barely hits the leading edge of the Earth and one where it barely hits the trailing edge of the Earth. The risk corridor spans between the leading and the trailing grazing cases.

Risk Corridor Calculation Validation

To validate the module, impact points for asteroid 2011 AG5 with a possible impact* on February 5th 2040 were computed. The result is shown in Fig. 3 and can be compared to the result obtained by Adamo[20] in Fig. 4. The shape and locations of the impact points coincide well based on visual inspection. Impact velocities are similar and only differ to about 0.1 km s$^{-1}$ in the regions where the data are available in Fig. 4. Impact angles are expected to decrease to 0° at both ends of the impact location line corresponding to near tangential impacts. This behaviour can be observed in Fig. 3. One data point is labelled with the impact angle in Fig. 4 and the angle agrees to about 5° with a similarly located point in Fig. 3. Overall, the results agree well with each other and suggest accurate performance of the Risk Corridor Calculation module.

---

* NASA's and ESA's Near Earth Object (NEO) risk lists are updated continuously. Thus, asteroids can always be removed or added to the list depending on newly made observations. This is why it is possible that some of the asteroids mentioned in this publication might be removed from the risk list after publishing. At the time of writing, 2011 AG5 was removed from NASA's and ESA's risk list because new observations ruled out the possibility of a future impact.

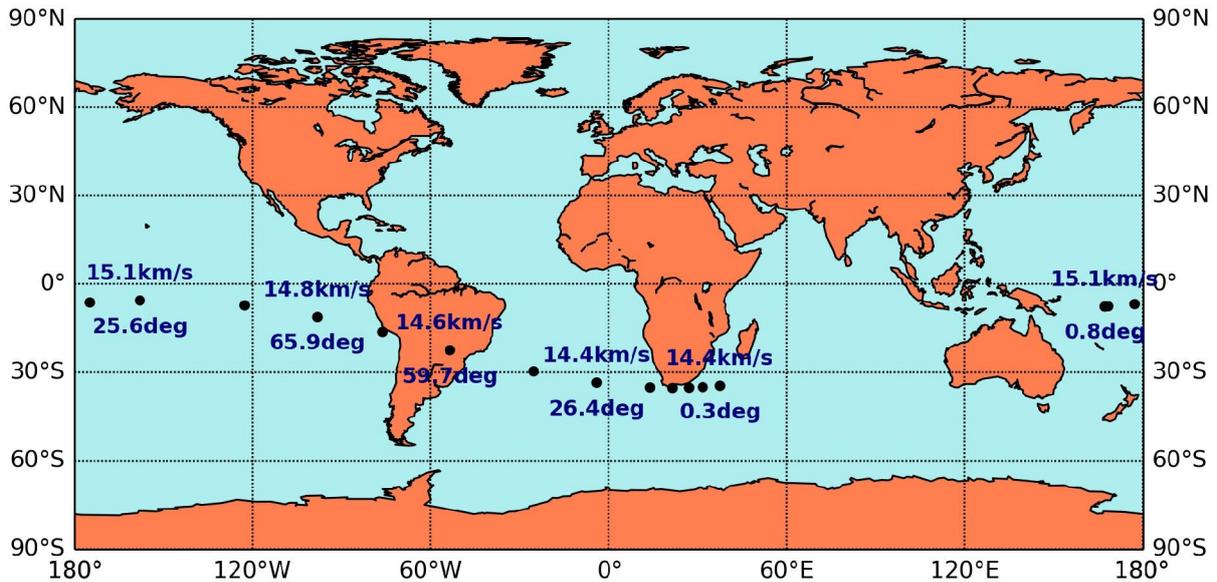

Fig. 3: Recorded impact points for 2011 AG5. Impact speed and angle are noted above and below the impact point, respectively. A zero degree impact angle corresponds to a tangential impact and marks a grazing case.

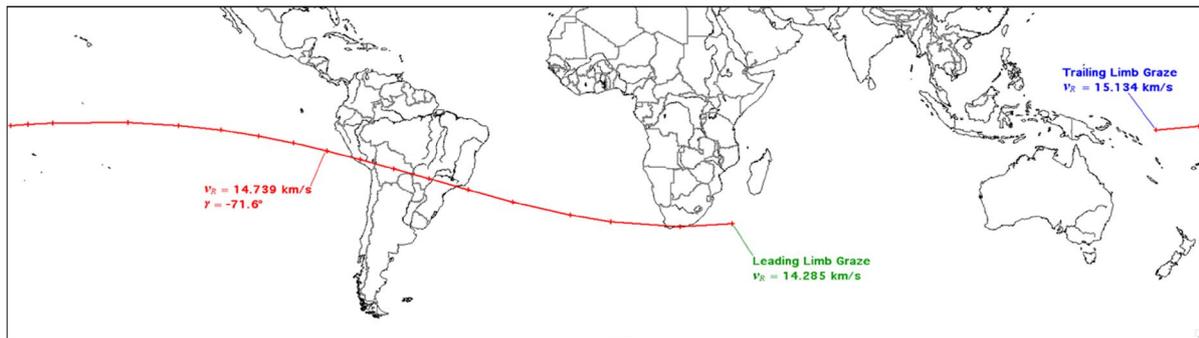

Fig. 4: Impact line for 2011 AG5 as published by Adamo[20]. The flight path angle in the middle of the corridor is marked as -71.6° and its norm corresponds to the impact angle.

Impact Location Probability Distribution

The recorded intermediate impact points were used to construct a spline in between the leading and trailing grazing case. This is the centre line of the impact location corridor. Because the orbit solution of a potentially impacting asteroid has an ellipsoidal uncertainty region associated with it, the impact point can deviate perpendicularly from the corridor centre line. It was assumed that a Gaussian distribution centred on the impact location centre line reflects adequately this condition. NASA's Near Earth Object (NEO) risk list provides the 1-σ width of the orbit solution uncertainty region for each asteroid. In this work, it was assumed, that this width maps directly onto the surface of the Earth. Thus, the Gaussian distribution centred on the impact location line was assigned the same standard deviation as the width of the orbit solution. This assumption was also suggested by Adamo[20]. Fig. 5 shows an example of a Gaussian distribution centred on the impact location centre line.

II. IV World Population Data

To gain some understanding of which regions on Earth are affected by an asteroid impact, world population data can be convolved with the impact distribution. The world population database can be accessed online[21] and a map showing the world population as estimated for the year 2015 is shown in Fig. 6.

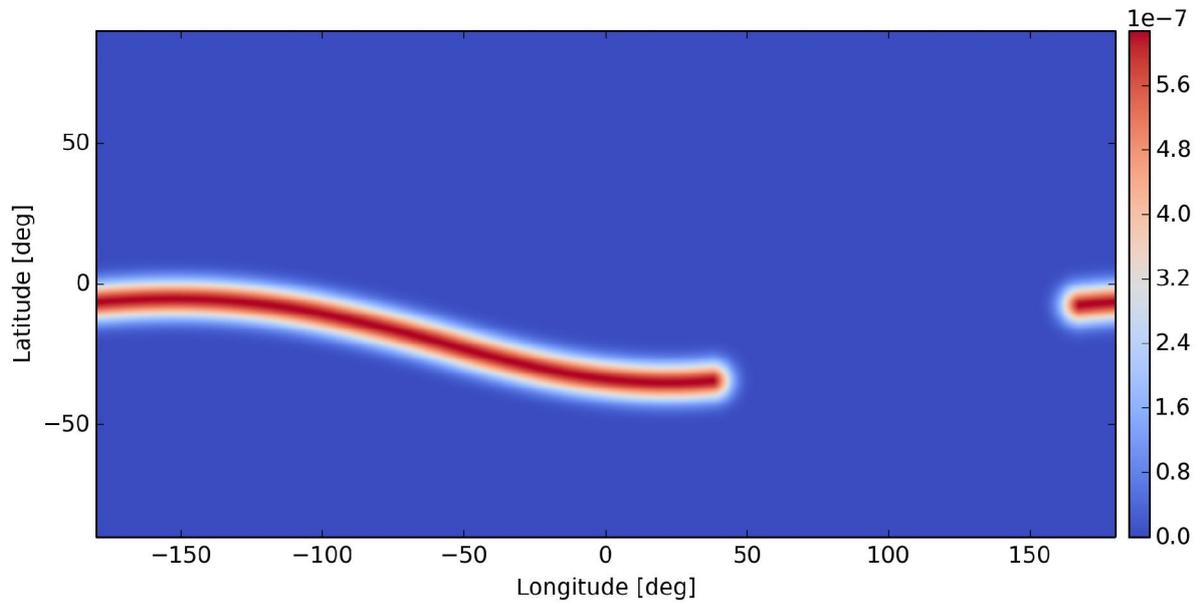

Fig. 5: Illustration of an impact probability distribution assuming a Gaussian distribution placed on the impact location centre line. The width has been exaggerated for visualisation purposes. A unitary accumulated impact probability is assigned to this map.

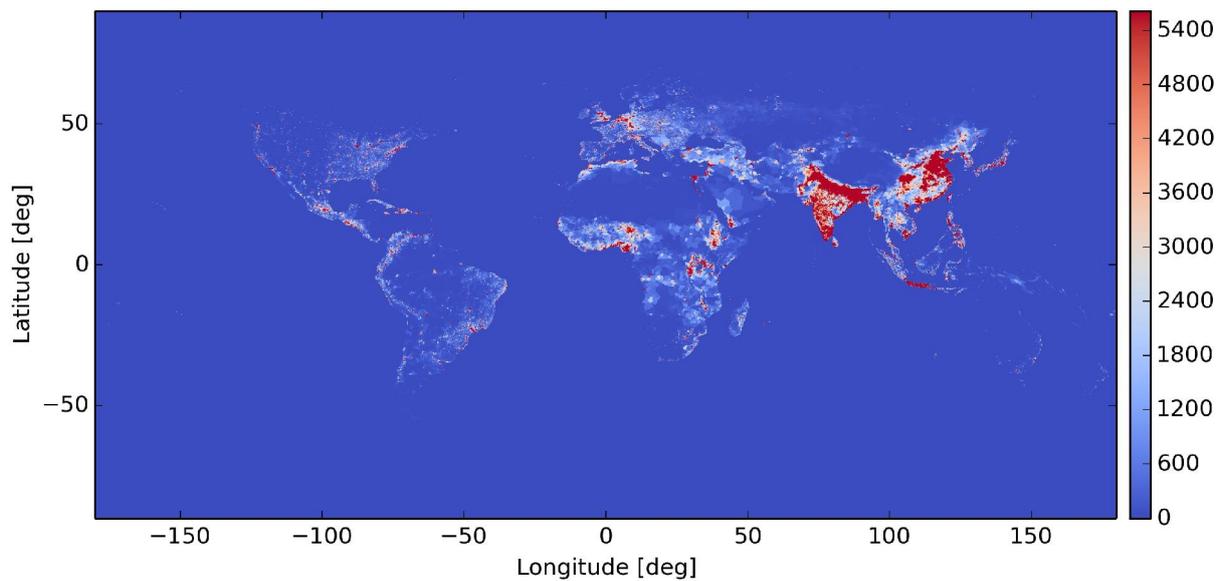

Fig. 6: Estimate of world population for the year 2015. The data[21] have a resolution of 2.5'x2.5'. The color-coding is truncated at a maximum population of 5400 persons per cell for visualisation purposes. The actual maximum number is higher.

II.V Risk calculation

Forecasted risk is the product of the probability that an event occurs and the expected loss in that event. Here, the potential event is an asteroid impact. The probability of that event becomes quantifiable with the impact location probability distribution. The expected loss of an impact event is the number of deaths in the population map grid cell (see Fig. 6) where the asteroid hits. The following paragraphs elaborate on the probability calculation and the impact effect assumptions.

International Astronautical Congress 2014 - "Move An Asteroid" competition winning paper

Probability Calculation

At this point, the accumulated impact probability is assumed to be unity on the world map with relatively high impact probabilities along the corridor centre line and insignificant impact probabilities at corridor-distant points. In other words, it is assumed that the asteroid will hit the Earth and the corridor provides information about the likely impact location. However, the asteroid usually has a small chance of impacting the Earth to begin with. This is why, for the risk calculation, the impact location probability distribution (example shown in Fig. 5) is multiplied with the global impact probability of that asteroid. The global impact probability for each asteroid is retrieved from NASA's NEO risk list. The result is an absolute impact probability value for each grid cell on the world map.

Impact Effect Assumption

For the risk calculation, it is assumed that the expected loss of an asteroid impact is the instant death of the population in the world map grid cell where the asteroid hits. The size of one grid cell is 2.5x2.5 arcmin$^2$ which is equivalent to about 5x5 km$^2$ at the equator. This statement is associated with significant simplifying assumptions about the impact effect as it accounts only for impact location and assumes that everyone who is present in that grid cell dies. It disregards asteroid properties such as size, composition and the impact velocity vector as well as the situation on the ground such as time of day and shelter strength. Furthermore, secondary impact effects, such as tsunamis, blast waves, crater ejecta and thermal radiation can affect population far beyond one grid cell. While it disregards many impact mechanisms of which some are moderating and others are amplifying, the assumption made here allows for a quick first estimate of impact mortality.

III. RESULTS

The methods described in the previous section were applied to 315 asteroids on NASA's NEO risk list. Given their ephemerides, an attempt was made to find their virtual impactors. If an impact could be detected, the impact location probability distribution was calculated. Finally, the population at risk was identified. Typically, this process took about five to ten minutes for each asteroid on a 3.4 GHz computer system.

III.I. Virtual Impactors

For ten of the 315 asteroids, an impact could be determined. This corresponds to a low yield rate of 3.2%. The impacting asteroids are listed in Table I. The table also lists the detected impact date, asteroid diameter, global impact probability and 1-σ corridor width for each of the asteroids.

| Name | Impact Date | Size [m] | Impact Probability | Corridor Width (1 σ) [$R_E$] |
|---|---|---|---|---|
| 2011AG5 | 2040/2/5 | 140 | 1.557E-3 | 1.17E-3 |
| 2009JF1 | 2022/5/6 | 13 | 3.5E-4 | 5.39E-4 |
| 2005BS1 | 2016/1/14 | 12 | 7.2E-5 | 8.82E-3 |
| 2011SO189 | 2056/9/24 | 17 | 1.2E-6 | 9.23E-3 |
| 2012DW60 | 2082/3/13[+] | 17 | 1.6E-5[+] | 1.0E-4 |
| 2010JH110 | 2070/6/2[+] | 19 | 4.1E-6[+] | 1.96E-3 |
| 2011SM173 | 2058/9/22 | 9 | 5.0E-5 | 2.11E-3 |
| 2009JL2 | 2057/5/13 | 22 | 1.0E-6 | 7.46E-4 |
| 2006WP1 | 2015/11/17 | 8 | 4.9E-7 | 1.62E-2 |
| 2008KT | 2089/11/19[+] | 8 | 4.6E-8[+] | 1.39E-3 |

Table I: List of impacting virtual asteroids. Impact probability and corridor width were obtained from NASA's NEO risk list. A superscript ([+]) means that an impact date different to that listed by NASA was discovered[†]. In this case, impact probability and the corridor width were taken from the nearest impact date.

For the majority of asteroids, no impact could be detected. Table II bins the majority of these cases according to their maximum miss distance up to 600 Earth radii.

| Miss distance | ≤ 6.5 $R_E$ (GEO) | ≤ 60 $R_E$ (Moon) | ≤ 600 $R_E$ | ≥600 $R_E$ |
|---|---|---|---|---|
| # | 45 | 83 | 72 | 68 |
| % | 14.3 | 26.3 | 22.9 | 21.6 |

Table II: Statistics about Earth-missing virtual asteroids. The total number is given as well as the percentage with respect to 315 analysed asteroids. $R_E$ is the radius of Earth (6378 km). The geostationary orbit is about 6.5 $R_E$ away, and the Moon about 60 $R_E$. Asteroids that could not be determined were excluded.

III.II. Impact Location Probability Distribution

The possible impact corridors for the asteroids listed in Table I were computed. Each (so far normalized to unitary impact risk) impact corridor was multiplied with its global impact probability as listed in Table I. Fig. 7 shows a collage of all ten corridors featuring their respective widths and probabilities in the world grid. Because the probabilities differ by orders of magnitude, the map is provided on a logarithmic (base ten) scale. The certainty of the orbit solution of each asteroid determines corridor width. It is apparent that some asteroids have a higher orbit certainty associated with them as their corridors are thinner. White regions are associated with zero impact probability for the ten analysed asteroids.

---

[†] All asteroids with differing impact dates exhibit constant periodicity in their potential impacts with a period of one year. In all cases, an impact was calculated to occur one or two years before the first date listed on NASA's NEO risk list. This means that the impact dates found here, match the general pattern of impact dates.

III.III. Global Asteroid Risk Map

The probability distributions shown in Fig. 7 were convoluted with global population data shown in Fig. 6 to obtain a global asteroid risk map. The result is shown in Fig. 8. The highest risk is associated with the corridor of 2011AG5 that stretches over South America. This section of the corridor spans over a relatively densely populated area and has a very high impact probability associated with it. As a result, it dominates the total impact risk. The total global risk associated with the ten analysed asteroids is the sum of all the map cells in Fig. 8 and evaluates to 29,919 people

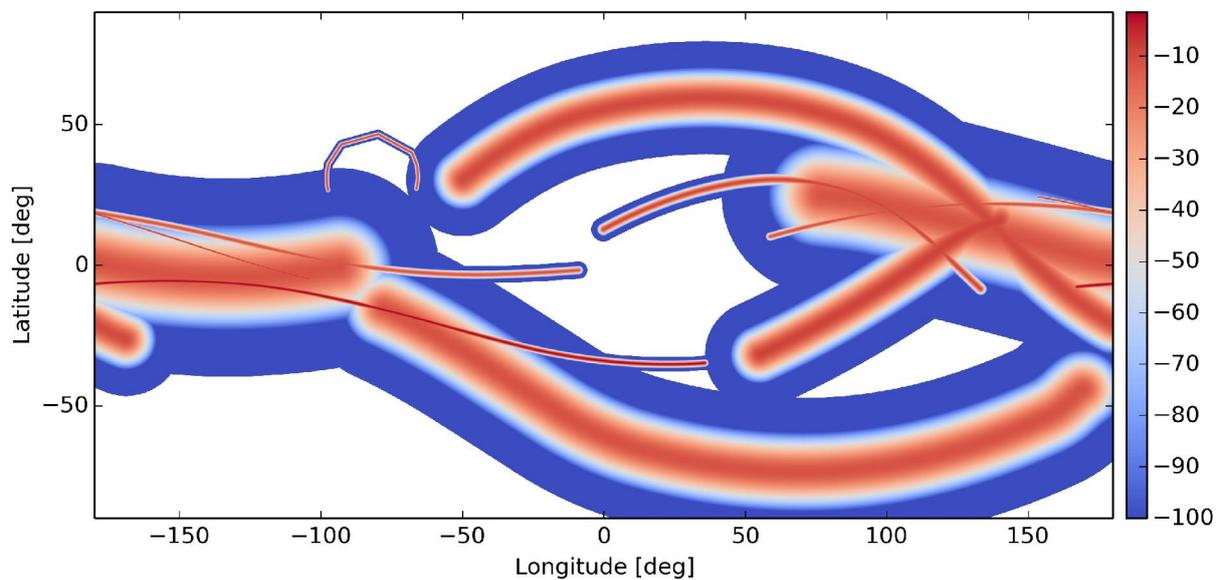

Fig. 7: Global impact location probability distributions of the ten asteroids listed in Table I. The map is colour coded on a base ten logarithmic scale.

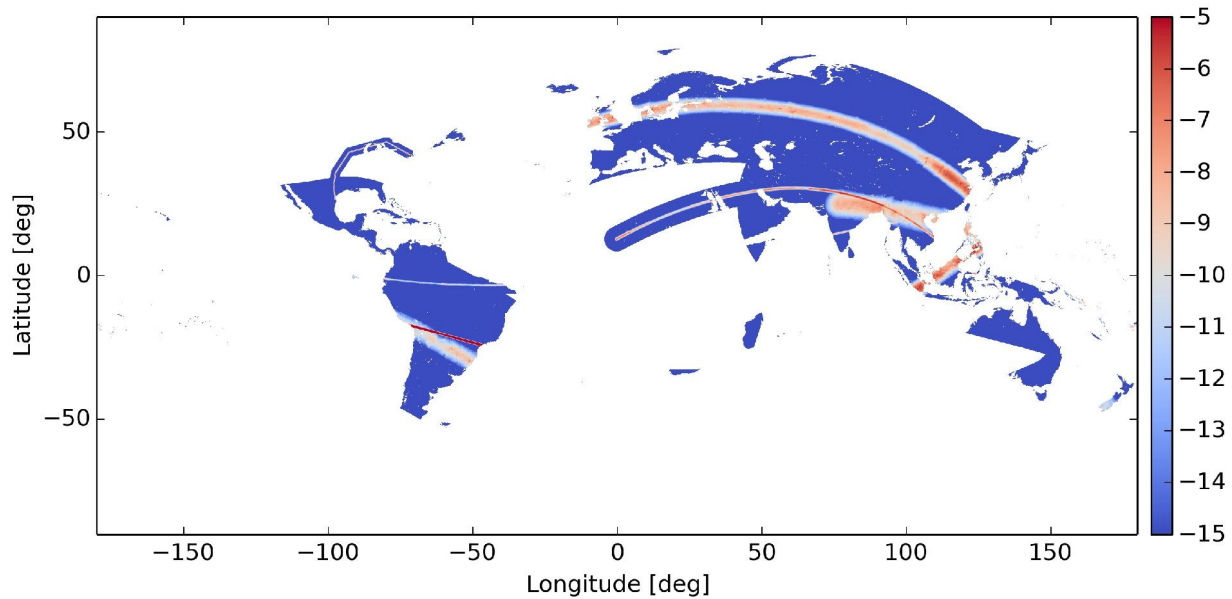

Fig. 8: Global asteroid risk map colour coded on a base ten logarithmic scale. White regions are associated with a risk that is effectively zero. The total risk evaluates to 29,919 casualties.

## IV. DISCUSSION

This section discusses the results presented in the previous section.

### IV.I. LOV Method

To find impacting virtual asteroids, the ephemeris belonging to the nominal orbit solution was varied along the LOV. Albeit a simple and fast way of searching for virtual impacting asteroids, the method failed to find virtual impactors of many asteroids that have a non-zero chance of impacting the Earth in the future according to the risk lists maintained by ESA and NASA. Only about 3% of the virtual impactors from the risk list could be found using the LOV method. This outcome suggests that the LOV method is not very efficient. The reason is that the search space is closely constricted. Only a minority of virtual impactors lies precisely on the one-dimensional LOV. However, the uncertainty region is a three-dimensional ellipsoid and it is therefore likely that a virtual impactor is offset from the LOV. Already small deviations from the LOV are sufficient for a virtual impactor to miss the Earth. This interpretation is corroborated by the fact that many "near-misses" were found as shown in Table II. 45 virtual asteroids passed the Earth within a distance equivalent to the geostationary orbit and another 82 virtual asteroids passed within a distance equivalent to the semi-major axis of the Moon. On a solar system scale, these cases are very close encounters. On the other hand, 72 asteroids passed within 600 Earth radii and 69 at even greater distances. These far misses are remarkable considering that the tested ephemerides are associated with potentially impacting asteroids. One explanation is that the search algorithm iterates the ephemeris towards the next local minimum starting at the nominal orbit solution. However, the next local minimum might not be very deep and can be on the opposite side on the LOV than the impact solution. In this case, the algorithm settles on a far miss distance. It is likely that many (possibly all) of these misses turn into impacts when the search space is not constricted to one line and employs a more sophisticated sampling method.

### IV.II. Unconfirmed Impact Dates

In Table I, three asteroids are associated with impact dates that are not listed on the official NEO risk lists. However, these three asteroids show a near constant periodicity of one year in their reoccurring impact dates. The impact dates determined by ARMOR are one or two years earlier than those listed by NASA. This fact, in conjunction with the regular periodicity, suggests that the asteroids have close flybys with the Earth prior to their first impact. It is therefore plausible that because of small differences in propagator performance, ARMOR detects an impact where NASA detects a close flyby.

### IV.III. Asteroid Risk

The asteroid risk is global in nature. Even with a small sample size of ten asteroids, the possible impact locations mapped in Fig. 8 draw a clear picture that every region on the Earth is potentially affected. For the ten asteroids that are analysed in this study, the total risk equates to 29,919 casualties.

Because the system does not take into account impact effects, the estimated risk value is not reliable on an absolute scale. Despite this limitation, the value serves well as a quick benchmark to compare the risks associated with specific asteroids. It also allows identification of high risk regions on the Earth.

In general, highly populated areas have a naturally higher risk because their expected loss in case of an impact is greater. It is therefore not surprising that regions in China, India and Southeast Asia (see Fig. 6) stand out on the asteroid global risk map in Fig. 8. However, risk is also dependent on impact probability and the asteroid 2011AG5 is associated with a relatively high global impact probability (see Table I). Furthermore, the impact corridor of this asteroid spans a densely populated region across South America including Sao Paolo. Densely populated impact location in conjunction with a high impact probability mean that asteroid 2011AG5 dominates the total impact risk.

It should be noted that the predictions made in association with the asteroid risk reach far into the future. However, risk associated with the asteroid threat is highly time-dependent. New observations may change the impact probability distribution. Furthermore, the situation on the ground changes over time. In this paper, potential impacts up to the year 2100 were considered. It is a considerable challenge to anticipate the population situation, and therefore the expected loss factor, in this timeframe. The population data used to compute the asteroid risk anticipates the world population situation in 2015. Because this is a short-term prediction, high confidence can be placed in the data. However, the population will vary significantly over the next 100 years[‡]. The used population data will lose its validity in the future and risk prediction will be associated with high uncertainty. For accurate risk estimation, the expected loss in the event of an impact needs to be known. Knowledge about the population situation at the time of impact is therefore a crucial component in risk calculation. If reliable future population data are inaccessible, expressing the number of people affected by an asteroid threat as a percentage of the world population is one way of moderating the effect of inaccurate population knowledge. This method makes the risk assessment robust against global population increases or decreases. However, if the spatial distribution of world population changes significantly, this approach would be ineffective.

### IV.IV. Future Application of ARMOR

This paper aimed to visualize and quantify the global asteroid risk. This is useful in its own right as it helps to characterize and make tangible a risk that is not prevalent in the public consciousness. Beyond its capability to address the global nature of the asteroid risk, the ARMOR tool may also be used to characterize the specific risk posed by asteroids of special interest. When, for example, a new potentially hazardous object (PHO) is discovered, ARMOR can be employed in conjunction with dedicated impact effect simulators to quantify accurately the impact risk.

## V. CONCLUSIONS

The ARMOR tool which is currently being developed maps asteroid ephemerides to impact probability distributions on the surface of the Earth. By convoluting the potential impact locations with Earth population data, global impact risk can be approximated. The concept and accurate performance of the system was demonstrated with ten asteroids. The modules of the system were positively validated against results found in the literature. Further research on certain aspects of the system promise considerable capability improvements in the future. A more sophisticated search algorithm has great potential to increase the virtual impactor search yield, and, thus add much value to the overall system. The search space of the algorithm needs to be generalized to find virtual impactors beyond the LOV. Impact effects are mostly disregarded and need further attention in the future. However, the proposed method was suitable for the current aim which was to convey the global nature of the asteroid threat. Even though the sample size of ten asteroids is limited, it is clear from the result that the asteroid threat is a global issue. The risk corridors shown in Fig. 8 are distributed over the entire globe.

Beyond its current capabilities, the ARMOR tool will be useful in the future to study the risk posed by individual asteroids of special interest. It can be utilized for decision making on the appropriate course of action in response to an asteroid threat – be it passive or active in nature.

## VI. ACKNOWLEDGEMENT

The author thanks his supervisors. Dr Hugh G Lewis and Professor Peter M. Atkinson at the University of Southampton who both contributed tremendously to the quality of this paper with their support, constructive criticism and engaging discussions.

---

[‡] For perspective: Population increased from 1.6 to 6.1 billion people in the period from 1900 to 2000.